\documentclass[12pt]{article}
{\baselineskip 12pt}  
\begin{document}
\title{
\vspace*{-30pt} Need the masses of unstable particles \\ and their
antiparticles be equal \\in the CPT--invariant system?}
\author{ \hfill \\ K. Urbanowski\footnote{e--mail:
K.Urbanowski@if.uz.zgora.pl; K.Urbanowski@proton.if.uz.zgora.pl} \\  \hfill  \\
University of Zielona Gora, Institute of Physics, \\
ul. Podgorna 50, 65-246 Zielona Gora, Poland. \vspace*{-10pt}}
\maketitle {\noindent}{\em PACS numbers:}  03.65.Ca., 11.30.Er.,
11.10.St., 14.40.Aq. 
\begin{abstract}
We show that the real parts of diagonal matrix elements of the
exact effective Hamiltonian governing the time evolution in the
subspace of states of neutral kaons and similar particles can not
be equal for $t > t_{0}$ ($t_{0}$ is the instant of creation of
the pair $K_{0}, {\overline{K}}_{0}$) when the total system under
consideration is CPT invariant but CP noninvariant. The unusual
consequence of this result is that, contrary to the properties of
stable particles, the masses of the unstable particle, e.g..
$K_{0}$, and its antiparticle, ${\overline{K}}_{0}$,  need not be
equal for $t \gg t_{0}$ in the case of preserved CPT and violated
CP symmetries.
\end{abstract}

\section{Introduction}

All known CP--  and hypothetically  possible CPT--violation
effects in the neutral kaon complex are  described by solving the
Schr\"{o}dinger--like evolution equation \cite{Lee1} ---
\cite{improved} (we use $\hbar = c = 1$ units)
\begin{equation}
i \frac{\partial}{\partial t} |\psi ; t >_{\parallel} =
H_{\parallel} |\psi ; t >_{\parallel}, \; \; \; (t \geq t_{0}),
\label{LOY}
\end{equation}
for $|\psi ; t >_{\parallel}$ belonging to the subspace ${\cal
H}_{\parallel} \subset {\cal H}$ (where ${\cal H}$ is the state
space of the physical system under investigation), e.g., spanned
by orthonormal neutral  kaons states $|K_{0}>, \;
|{\overline{K}}_{0}>$, and so on, (then states corresponding with
the decay products belong to ${\cal H} \ominus {\cal
H}_{\parallel} \stackrel{\rm def}{=} {\cal H}_{\perp}$), and
nonhermitian effective Hamiltonian $H_{\parallel}$ obtained
usually by means of  the Lee--Oehme--Yang (LOY) approach (within
the Weisskopf--Wigner approximation (WW)) \cite{Lee1} ---
\cite{improved}:
\begin{equation}
H_{\parallel} \equiv M - \frac{i}{2} \, \Gamma, \label{H-eff}
\end{equation}
where $M = M^{+}, \; \Gamma = {\Gamma}^{+}$ are $(2 \times 2)$
matrices. In a general case $H_{||}$ can depend on time $t$,
$H_{||} \equiv H_{||}(t)$, \cite{horwitz,acta83,11}. The effective
Hamiltonians $H_{||}$ are usually derived by rewriting  the
Sch\"{o}dinger equation in the kaon rest frame:
\begin{equation}
i \frac{\partial}{\partial t} U(t)|{\psi} >_{||} = H
U(t)|{\psi}>_{||}, \; \; U(t = t_{0}) = I, \label{l2}
\end{equation}
where $I$ is the unit operator in $\cal H$, $H$ is be total
(selfadjoint) Hamiltonian acting in $\cal H$ and $U(t)$  the total
unitary evolution  operator, $|{\psi}>_{||}\, \equiv $ \linebreak
$|{\psi},t = t_{0}> _{||}\, \in \, {\cal H}_{||}$ is the initial
state of the system.

Solutions of Eq. (\ref{LOY}) can be written in matrix  form  and
such a  matrix  defines  the evolution operator (which    is
usually nonunitary) $U_{\parallel}(t)$ acting in ${\cal
H}_{\parallel}$:
\begin{equation}
|\psi ; t >_{\parallel}  \stackrel{\rm def}{=} U_{\parallel}(t)
|\psi >_{\parallel}, \label{l1a}
\end{equation}
where,
\begin{equation}
|\psi >_{\parallel} \equiv q_{1}|{\bf 1}> + q_{2}|{\bf 2}>,
\label{l1b}
\end{equation}
and $|{\bf 1}>$ stands for  the  vectors of the   $|K_{0}>,  \;
|B_{0}>$ type and $|{\bf 2}>$ denotes antiparticles  of  the
particle "1": $|{\overline{K}}_{0}>, \; {\overline{B}}_{0}>$,
$<{\bf j}|{\bf k}> = {\delta}_{jk}$, $j,k =1,2$.

Relations between matrix elements of  $H_{\parallel}$  implied  by
the CPT--trans\-for\-ma\-tion properties of the Hamiltonian $H$ of
the total system, containing the  neutral  kaon  complex  as a
subsystem,  are crucial for designing CPT--invariance and
CP--violation tests  and for proper interpretation of their
results. The standard interpretation of matrix elements, $h_{jk} =
<{\bf j}|H_{||}|{\bf k}>$, ($j.k = 1,2$),  of the effective
Hamiltonian $H_{||}$ follows form the properties of the LOY
effective Hamiltonian $H_{||} \equiv H_{LOY}$. In many papers it
is assumed that the real parts, $\Re (.)$, of the diagonal matrix
elements of $H_{\parallel}$:
\begin{equation}
\Re \, (h_{jj} ) \equiv M_{jj}, \; \;(j =1,2), \label{m-jj}
\end{equation}
where
\begin{equation}
h_{jk}  =  <{\bf j}|H_{\parallel}|{\bf k}>, \; (j,k=1,2),
\label{h-jk}
\end{equation}
correspond to the masses  of particle "1" and its antiparticle "2"
respectively \cite{Lee1} --- \cite{improved}, (and such an
interpretation of $\Re \, (h_{11})$ and $\Re \, (h_{22})$ will be
used in this paper), whereas the imaginary parts, $\Im (.)$, are
interpreted as the decay widths of these particles, ${\Gamma}_{jj}
\equiv -2 \Im \,(h_{jj})$, $(j =1,2)$, \cite{Lee1}
--- \cite{improved}, \cite{data}.

Taking $H_{||} = H_{LOY}$ and assuming that the CPT invariance
holds in the system considered one easily finds the standard
result of the LOY approach
\begin{equation}
h_{11}^{LOY}  = h_{22}^{LOY}, \label{LOY-h=h}
\end{equation}
which, among others, means that
\begin{equation}
M_{11}^{LOY} = M_{22}^{LOY}, \label{LOY-m=m}
\end{equation}
where $M_{jj}^{LOY} = \Re \, (h_{jj}^{LOY})$ and $h_{jj}^{LOY} =
<{\bf j}|H_{LOY}|{\bf j}>$, ($j =1,2$). This last relation is
interpreted as the equality of masses of the unstable particle
$|{\bf 1}>$ and its antiparticle $|{\bf 2}>$.

In \cite{plb2002} using the Khalfin's Theorem \cite{chiu,leonid1}
--- \cite{kabir}  it was proved that, contrary to the conclusion drawn in
(\ref{LOY-h=h}), the diagonal matrix elements of the exact
effective Hamiltonian $H_{||}$ in a CPT invariant but CP
noninvariant system  must be different for $t > t_{0}$. This proof
is rigorous. A crucial conclusion following from this property was
put forward in \cite{plb2002}. This conclusion states that,
contrary to property (\ref{LOY-m=m}),  the masses of unstable
particle and its antiparticle should be different for $t> t_{0}$.
In fact, this conclusion was not supported there by a direct and
rigorous proof. The aim of this note is to complete the result of
\cite{plb2002} and to prove this conclusion. Strictly speaking, to
prove rigorously that in the case of the exact $H_{||}$ it must be
$\Re \,(h_{11} - h_{22}) \, \neq \, 0$ for $t
> t_{0}$, if the CPT symmetry holds and CP is violated. In
order to realize this purpose, the method applied in \cite{11}
will be used.

\section{CPT transformation properties \\of the exact $H_{\parallel}$ }

The aim of this Section is to show that in the case of the exact
$H_{||}$, the standard LOY relation (\ref{LOY-m=m}) does not occur
if  the total system under consideration is CPT invariant,
${\Theta}H{\Theta}^{-1} = {H}^{+} \equiv H$ , that is
\begin{equation}
[ \Theta , H ] = 0, \label{[CPT,H]}
\end{equation}
where $\Theta$ is the antiunitary operator:
\begin{equation}
\Theta \stackrel{\rm def}{=} {\cal C}{\cal P}{\cal T},
\label{new4}
\end{equation}
and CP noninvariant.

Let $P$ denote the projection operator onto the subspace ${\cal
H}_{\parallel}$:
\begin{equation}
P{\cal H} = {\cal H}_{\parallel}, \; \; \; P = P^{2} = P^{+},
\label{new2}
\end{equation}
then the subspace of decay products ${\cal H}_{\perp}$ equals
\begin{equation}
{\cal H}_{\perp}  = (I - P) {\cal H} \stackrel{\rm def}{=} Q {\cal
H}, \; \; \; Q \equiv I - P. \label{l7c}
\end{equation}
For the  case of neutral  kaons  or  neutral  $B$--mesons,  etc.,
the projector $P$ can be chosen as follows:
\begin{equation}
P \equiv |{\bf 1}><{\bf 1}| + |{\bf 2}><{\bf 2}|. \label{l3}
\end{equation}
We assume that time independent basis vectors $|K_{0}>$ and
$|{\overline{K}}_{0}>$ are defined analogously to corresponding
vectors used in the LOY theory of time evolution in the neutral
kaon complex \cite{Lee1} --- \cite{improved}: Vectors $|K_{0}>$
and $|{\overline{K}}_{0}>$ can be identified with the eigenvectors
of the so--called free Hamiltonian $H^{(0)} \equiv H_{strong} = H
- H^{(1)}$, where $H^{(1)} \equiv H_{int}$ denotes weak and other
interactions which are responsible for transitions between
eigenvectors of $H^{(0)}$, i.e., for the decay process. This means
that
\begin{equation}
[P, H^{(0)}] = 0. \label{new3}
\end{equation}
The condition guaranteeing the occurrence of transitions  between
subspaces ${\cal H}_{\parallel}$ and ${\cal H}_{\perp}$, i.e.,  a
decay process of states in ${\cal H}_{\parallel}$, can  be written
as follows
\begin{equation}
[P,H] \neq 0, \label{l7}
\end{equation}

Note that Eq. (\ref{l2}) means that $U(t) = \exp (-itH)$. Now,
knowing $U(t)$, the exact evolution  operator  $U_{\parallel}(t)$
(\ref{l1a}) for ${\cal H}_{\parallel}$ can be expressed  using the
projector $P$  as follows
\begin{equation}
U_{\parallel}(t) \equiv PU(t)P. \label{l4}
\end{equation}
We  have  $U_{\parallel}(0) \equiv  P$. In
\cite{horwitz,acta83,11} an observation was made that for every
effective Hamiltonian $H_{\parallel}$ governing  the time
evolution in subspace ${\cal H}_{\parallel}$ $\equiv P {\cal H}$,
the following identity holds:
\begin{equation}
H_{\parallel} \equiv H_{\parallel}(t)  \equiv i
\frac{{\partial}U_{\parallel}(t)}{\partial t}
[U_{\parallel}(t)]^{-1}, \label{l5}
\end{equation}
where the operator $[U_{\parallel}(t)]^{-1} $ is defined as
follows \cite{plb2002}
\begin{equation}
[U_{\parallel}(t)]^{-1} U_{\parallel}(t) = U_{\parallel}(t)
[U_{\parallel}(t)]^{-1}\equiv P. \label{l6}
\end{equation}
(These last two relations were also used in \cite{plb2002}). It
can be easily verified  that  $H_{\parallel}  \equiv  H_{LOY}$,
fulfills the identity (\ref{l5}).

In the nontrivial case (\ref{l7}) from (\ref{l5}) (see Appendix,
formula  (\ref{A2})), using (\ref{l2}) and (\ref{l4}) we find
\begin{eqnarray}
H_{\parallel}(t) & \equiv &  PHU(t)P[U_{\parallel}(t)]^{-1} P
\label{l7a1}  \\
& \equiv & PHP + PHQ U(t) [ U_{\parallel}(t)]^{-1}P \label{l7a} \\
& \stackrel{\rm def}{=} & PHP + V_{\parallel}(t).  \label{l7b}
\end{eqnarray}
Thus \cite{8,9,10,pra}
\begin{equation}
H_{\parallel}(0) \equiv PHP, \; \; V_{\parallel}(0) = 0, \; \;
V_{\parallel} (t \rightarrow 0) \simeq -itPHQHP, \label{l8}
\end{equation}
so,   in   general   $H_{\parallel}(0)    \neq$ $H_{\parallel}(t
\gg t_{0}=0)$ \cite{8,9,10} and $V_{\parallel}(t \neq 0) \neq
V_{\parallel}^{+}(t \neq 0)$, $H_{\parallel}(t \neq 0) \neq
H_{\parallel}^{+}(t \neq 0)$.

Now let  us  pass  on  to  the investigation  of the
CPT--transformation properties of $H_{\parallel}$. We assume that
vectors $|{\bf 1}>,|{\bf 2}>$ are related to each other through
the transformation:
\begin{equation}
\Theta |{\bf 1}> \stackrel{\rm def}{=} - |{\bf 2}>, \; \; \Theta
|{\bf 2}> \stackrel{\rm def}{=} - |{\bf 1}>, \label{CPT1=2}
\end{equation}
Besides, there is only one assumption for the anti--linear
operator ${\Theta}$ (\ref{new4}) describing the
CPT--transformation in $\cal H$. We require CPT--invariance of
${\cal H}_{\parallel}$. This means that for the projector $P$
defining this subspace the following relation must hold,
\begin{equation}[ \Theta ,P] = 0.
\label{[CPT,P]}
\end{equation}

Using assumption (\ref{[CPT,P]}) and  the   identity (\ref{l7a1}),
after some algebra, one finds \cite{hep-ph/9803376} (see Appendix
A)
\begin{equation}
[\Theta, H_{\parallel}(t)] = {\cal A}(t) + {\cal B}(t),
\label{l10}
\end{equation}
where:
\begin{eqnarray}
{\cal A}(t) \; & = & \; P  [{\Theta},H] U(t) P
\bigl( U_{\parallel}(t) {\bigr)}^{-1} , \label{l11} \\
{\cal B}(t) \; & = & \;  \Big{\{} PH  -  PH U(t)  P \bigl(
U_{\parallel}(t) {\bigr)}^{-1}P \Big{\}} [{\Theta} ,U(t)] P \bigl(
U_{\parallel}(t) {\bigr)}^{-1}
\label{l12}  \\
& \equiv &  { \Big{\{} } PH  \: -  \:  H_{\parallel}(t) P {
\Big{\}} } [{\Theta} ,U(t)]  P \bigl(
U_{\parallel}(t){\bigr)}^{-1}
\label{l13} \\
& \equiv &  { \Big{\{} } PHQ  \: -  \:  V_{\parallel}(t) P {
\Big{\}} } [{\Theta} ,U(t)]  P \bigl(
U_{\parallel}(t){\bigr)}^{-1} \label{l14}
\end{eqnarray}

We observe that for $t = 0$
\begin{equation}
{\cal A}(0) \equiv P[\Theta,H]P \; \; {\rm and} \; \; {\cal B}(0)
\equiv 0. \label{A=0}
\end{equation}
From definitions and the general properties of operators $\cal
C$,$\cal P$ and $\cal T$ \cite{5,6,messiah,bohm,15} it is known
that ${\cal T}U(t{\neq}0) = U^{+}(t{\neq}0){\cal T}$ $\neq$
$U(t{\neq}0){\cal T}$ (Wigner's definition for $\cal T$ is used),
and thereby ${\Theta}U(t \neq 0) = U^{+}(t \neq 0){\Theta}$
\cite{messiah,bohm,15} i.e. $[\Theta,U(t  \neq  0)] \neq  0$. So,
the   component ${\cal B}(t)$ in (\ref{l10}) is nonzero for $t
\neq 0$ and  it  is obvious that there is a chance for the
$\Theta$--operator to commute with the effective Hamiltonian
$H_{\parallel}(t   \neq     0)$ only if  $[\Theta,H] \neq 0$. On
the  other  hand,  the  property $[\Theta,H]\neq 0$ does not imply
that $[\Theta, H_{\parallel}(0)] =  0$  or $[\Theta,
H_{\parallel}(0)] \neq 0$. These two possibilities  are
admissible, but if $[\Theta, H] =  0$  then there  is  only  one
possibility: $[\Theta, H_{\parallel}(0)] = 0$ \cite{11}.

From (\ref{l10}) we find
\begin{equation}
\Theta H_{\parallel}(t) \Theta^{-1} - H_{\parallel}(t) \equiv
\bigl( {\cal A}(t) + {\cal B}(t) \bigr) \Theta^{-1}. \label{l16}
\end{equation}
Now, keeping in mind that $|{\bf 2}> \equiv  |\overline{K}_{0}>$
is the antiparticle for $|{\bf 1}> \equiv |K_{0}>$ and that,  by
definition, the  (anti--unitary) $\Theta$--operator  transforms
$|{\bf 1}>$   in  $|{\bf 2}>$   \cite{Lee2} --- \cite{6} according
to formulae (\ref{CPT1=2}), and $< \psi | \varphi > =
<{\Theta}{\varphi}|{\Theta}{\psi}>$, we obtain from (\ref{l16})
(see Appendix A)
\begin{equation}
h_{11}(t)^{\ast} - h_{22}(t) = <{\bf 2}| \bigl( {\cal A}(t) +
{\cal B}(t) \bigr) {\Theta}^{-1}|{\bf 2}>, \label{l17}
\end{equation}
Adding expression (\ref{l17}) to its  complex  conjugate  one gets
\begin{equation}
{\Re} \; (h_{11}(t) - h_{22}(t)) = {\Re} \; <{\bf 2}| \bigl( {\cal
A}(t) + {\cal B}(t) \bigr) {\Theta}^{-1}|{\bf 2}>. \label{l18}
\end{equation}

Note that if the requirement (\ref{l7}) for the projector $P$
(\ref{l3}) is replaced by the following one:
\begin{equation}
[P,H] = 0, \label{dd1}
\end{equation}
i.e. if only stationary states are considered instead of unstable
states, then one immediately obtains from (\ref{l11})
--- (\ref{l14}):
\begin{eqnarray}
{\cal A}(t) \; & = & \; P  [{\Theta},H] P,  \label{dd2} \\
{\cal B}(t) \; & = & \; 0. \label{dd3}
\end{eqnarray}

Let us assume that conditions (\ref{l7}) and (\ref{[CPT,H]}) hold.
For the stationary states (\ref{dd1}), the assumption
(\ref{[CPT,H]}), relations (\ref{dd2}), (\ref{dd3}) and
(\ref{l18}) yield ${\Re} \; (h_{11}(t) - h_{22}(t)) = 0$.

Now let us consider the case of unstable states, i.e., states
$|{\bf 1}>, |{\bf 2}>$, which lead to such projection operator $P$
(\ref{l3}) that condition (\ref{l7}) holds. If in this case
(\ref{[CPT,H]}) also holds then ${\cal A}(t=0) \equiv 0$ (see
(\ref{A=0})) and thus $[ {\Theta}, H_{\parallel}(0) ]$ $ = 0$,
which is in agreement with   earlier,  similar  results
\cite{11,plb2002}. This means that at $t = 0$:
\begin{equation}
{\Re} \; (h_{11}(0) - h_{22}(0)) = H_{11} - H_{22} = 0,
\label{H=0}
\end{equation}
where
\begin{equation}
H_{jk} = <{\bf j}|H|{\bf k}>, \; \; (j,k =1,2). \label{H-jk}
\end{equation}
Let $t > 0$. In this case  we have ${\Theta}U(t)  =
U^{+}(t)\Theta$, which gives ${\Theta}U_{\parallel}(t) =$
$U_{\parallel}^{+}(t)\Theta$, ${\Theta}U_{\parallel}^{-1}(t)  =
(U_{\parallel}^{+}(t))^{-1}\Theta$, and
\begin{equation}
[\Theta, U(t)]  =  - 2i \bigl( {\Im} \; U(t) \bigr) \Theta
\label{l20}
\end{equation}
This relation leads to the following result in  the  case  of
conserved CPT--symmetry
\begin{eqnarray}
{\cal B}(t) & = & -2i P \Big{\{}  H \: - \: H_{\parallel}(t) \: P
\Big{\}} \bigl( {\Im} \: U(t) \bigr) P \bigl( U_{\parallel}^{+}(t)
{\bigr)}^{-1}\Theta \label{l21}\\
& \equiv & -2i \Big{\{} PHQ \: - \: V_{\parallel}(t) \: P \Big{\}}
\bigl( {\Im} \: U(t) \bigr) P \bigl( U_{\parallel}^{+}(t)
{\bigr)}^{-1} \Theta , \label{l22}
\end{eqnarray}
which means that generally, in any case ${\cal B}(t > 0) \neq 0$.

Formulae (\ref{l21}), (\ref{l22}) allow us to conclude that $<{\bf
2}|{\cal B}(0){\Theta}^{-1} |{\bf 2}> = 0$ and ${\Re} \, <{\bf
2}|{\cal B}(t> 0) {\Theta}^{-1}|{\bf 2}> \neq 0$, if condition
(\ref{[CPT,H]}) holds. This means that in this case it must be
${\Re}\,(\,h_{11}(t)\,) \neq {\Re}\,(\,h_{22}(t)\,)$ for $t > 0$.
So, there is no possibility for ${\Re}\,(h_{11})$ to equal
${\Re}\,(h_{22})$ for $t> 0$  in  the  considered case of $P$
fulfilling the condition (\ref{l7}) (i.e., for unstable states)
when CPT--symmetry  is conserved.

Using identity (\ref{l7a1}) and assuming that $[{\cal C \cal P} ,
H ]= 0 $ and (\ref{[CPT,H]}) hold it is not difficult to show that
in such case $h_{11}(t) = h_{22}(t)$.

\section{Discussion.}

All  the above
considerations lead to the following conclusions  for  the matrix
elements $h_{jk}$   of   the exact effective    Hamiltonian
$H_{\parallel}$ governing the time evolution  in  neutral  kaons
subspace:\\
\hfill \\
{\it Conclusion 1}

If $[\Theta,H] = 0$ and $[{\cal C \cal P}, H] \neq 0$ then it
follows that $M_{11} \equiv {\Re \,}(\,h_{11}(t>0)\,)$ $\neq$  $
{\Re \,}(\,h_{22}(t>0)\,) \equiv M_{22}$, that is that the mass of
the unstable particle "1" must be different from the mass of its
antiparticle "2" for $t > t_{0} = 0$. \\
\hfill \\

One should remember that the above conclusion derived from
relation (\ref{l18}) concerns only the real parts of $h_{11}(t>0)$
and $h_{22}(t>0)$ and it is in excellent agreement with the
results presented in \cite{plb2002}.  Relations (\ref{l16})
--- (\ref{l18}) give us no information about the imaginary parts
of $h_{11}$  and $h_{22}$. One cannot infer from  ({\ref{l18})
that $[\Theta  ,H] =  0$ follows ${\Im}\,(h_{11}) \neq
{\Im}\,(h_{22})$. The case when $[ \Theta ,H] =0$ follows
${\Re}\,(\,h_{11}(t>0)\,) \neq {\Re} \, (\,h_{22}(t>0)\,)$ and
${\Im}\,(h_{11}) = {\Im}\,(h_{22})$,  is not in conflict with
relations (\ref{l16})
--- (\ref{l18}). The equality of $\Im \, (h_{11})$ and $\Im
\,(h_{22})$ need not imply the equality of $\Re \, (h_{11})$ and
$\Re \,(h_{22})$  and  vice versa. This means that the
Bell--Steinberger  relations  \cite{16} do not contradict
relations (\ref{l16}) --- (\ref{l18})  and {\it Conclusion 1}
following from them. Namely, Bell  and Steinberger formulae lead
to  the equality  of $\Im \, (h_{11})$  and $\Im \, (h_{22})$ in
the case of  conserved  CPT--symmetry and do not concern the real
parts of the diagonal matrix elements of $H_{\parallel}$  or give
any relations between them.

The real parts of the diagonal matrix elements of the mass matrix
$H_{\parallel}$, $h_{11}$ and $h_{22}$, are considered in the
literature as masses of unstable particles $|{\bf 1}>, |{\bf 2}>$
(e.g., mesons ${K}_{0}$ and ${\overline{K}}_{0}$). Such an
interpretation follows also from  properties, (\ref{l5}),
(\ref{l7b}) and (\ref{l8}) of the exact $H_{||}(t)$. The
interpretation of the diagonal matrix elements of $H_{||}(t=0)$ is
obvious (see (\ref{l8})). They have the dimension of the energy
(that is, the mass) and $h_{jj}(0) \equiv <{\bf j}|H|{\bf j}>$,
($j = 1,2$). So their interpretation as the masses of the particle
"1" and its antiparticle "2" at the instant $t = 0$ seems to be
explained. Note that from the identity (\ref{l5}) it follows that
the exact effective Hamiltonian $H_{||}(t)$ is a continuous
function of time $t$. Therefore the dimension of $H_{||}(t)$ as
the physical quantity at any $t > 0$ continues to be the same as
that at $t=0$.

From (\ref{l7b}) one finds
\begin{equation}
h_{jk}(t) = H_{jk} + v_{jk}(t), \;\;\;\; (j,k =1,2), \label{v-jk}
\end{equation}
where $v_{jk} (t) = <{\bf j}|V_{||}(t)|{\bf k}>$. So, at $t >
t_{0} = 0$ the initial mass (energy) of the particle $j$, $\Re \,
(h_{jj}(0)) = <{\bf j}|H|{\bf j}> \equiv H_{jj}$, in the state
$|{\bf j}>$,  shifts and takes the value
\begin{equation}
\Re \, (h_{jj}(t)) = H_{jj} + \Re \, (v_{jj}(t)). \label{Re-v-jj}
\end{equation}
Every experiment performed at this instant $t$ will indicate the
quantity (\ref{Re-v-jj}) as the energy (i.e., as the mass) of the
particle $j$ at time $t$. In the case of neutral particles there
are no methods allowing one to differentiate the contribution of
$\Re \, (h_{jj}(0)) = H_{jj}$ into $\Re \, (h_{jj}(t))$ from the
contribution of the shift $\Re \, (v_{jj}(t))$ by means of the
measurement performed at the instant $t > 0$. The particle $j$
always interacts with the environment at the instant $t$  as the
particle with the energy $\Re \, (h_{jj}(t))$. So from the point
of view of the relativistic quantum theory the interpretation of
$\Re \, (h_{jj}(t))$ as the mass of the particle $j$ seems to be
acceptable.

There is another one reason for the adoption of LOY interpretation
of the matrix elements of $H_{||}$ in this letter. Note that, as
it was mentioned in Sec. 2, the LOY effective Hamiltonian,
$H_{LOY}$ fulfils the identity (\ref{l5}). It seems that the
interpretation of matrix elements of any effective Hamiltonian
fulfilling this identity can not differ from the interpretation of
matrix elements of $H_{LOY}$.

So {\em Conclusion 1} means that masses of a decaying particle "1"
and its antiparticle "2" should be different if the CPT--symmetry
is conserved in the system containing these unstable particles. In
other words, in the exact theory unstable states $|{\bf 1}>, |{\bf
2}>$ appear to be nondegenerate in mass for $t > t_{0}$ if the
CPT--symmetry holds and the CP--symmetry does not, in the total
system considered.  At the same time, relations (\ref{dd1})
--- (\ref{dd3}) and (\ref{[CPT,H]}) suggest that in the CPT--invariant
system masses of a
given particle and its aniparticle are equal (i.e., appear to be
degenerate) only in the case of stationary (stable) states $|{\bf
1}>, |{\bf 2}>$. The case, when vectors $|{\bf 1}>, |{\bf 2}>$
describe pairs of particles $p, \overline{p}$, or $e^{-}, e^{+}$,
can be considered as an example of such states. All these
conclusions contradict the standard result of the LOY and related
approaches.

Results of the previous Section and {\em Conclusions 1} are not in
conflict with such implications of the CPT--invariance as the
equality of particle and antiparticle decay rates --- see
\cite{plb2002}. On the other hand the consequences (\ref{LOY-h=h})
and (\ref{LOY-m=m}) of the LOY theory are in conflict with the
results of Sec. 2  and {\em Conclusion 1} obtained without
approximations but they are in agreement with the rigorous result
obtained in \cite{plb2002}.

Note that in fact the above conclusions about the masses of
unstable particles under consideration are not in conflict with
the rigorous and consistent treatment of quantum theory. From
(\ref{[CPT,H]}) (or from the CPT Theorem \cite{cpt}) it only
follows that the masses of particle and antiparticle eigenstates
of $H$ (i.e., masses of stationary states for $H$) should be the
same in the CPT invariant system --- see \cite{plb2002}. Such a
conclusion can not be derived from (\ref{[CPT,H]}) for particle
$|{\bf 1}>$ and its antiparticle $|{\bf 2}>$ if they are unstable,
i.e., if states $|{\bf 1}>, |{\bf 2}>$ are not eigenstates of $H$.
Note also that the proof of the CPT Theorem makes use of the
properties of asymptotic states \cite{cpt}. Such states do not
exist for unstable particles. What is more, one should remember
that the CPT Theorem of axiomatic quantum field theory has been
proved for quantum fields corresponding to stable quantum objects
and only such fields are considered in the axiomatic quantum field
theory. There is no axiomatic quantum field theory of unstable
quantum particles. So, all implications of the CPT Theorem
(including those obtained within the S--matrix method) need not be
valid for decaying particles prepared at some initial instant
$t_{0} = 0$ and then evolving in time $t \geq 0$. Simply, the
consequences of CPT invariance need not be the same for systems in
which time $t$ varies from $t = - \infty$ to $t = + \infty$ and
for systems in which $t$ can vary only from $t = t_{0} > - \infty$
to $t = + \infty$. Similar doubts about the fundamental nature of
the CPT Theorem were formulated in \cite{kobayashi}, where the
applicability of this theorem for QCD was considered. One should
also remember that  conclusions about the equality of masses of
stable particles and their antiparticles following from the
properties of the S-matrix can not be extrapolated to the case of
unstable states. Simply, there is no S--matrix for unstable
states.

The important consequence of {\em Conclusion 1} is that the
conventional interpretation of the tests, which are sensitive to
the difference $\Re \, (h_{11} - h_{22})$, as the CPT invariance
test in neutral kaon complex, need not be longer valid. An example
of a such test is considered in \cite{hep-ph/0202253}.

Another consequence of the main result of the Section 2, that is
of the {\em Conclusion 1} concerns properties of the scalar
product of eigenvectors $|l>$, $|s>$ of $H_{||}$,
\begin{equation}
H_{||} |l(s)> \;= \; {\mu}_{l(s)}|l(s)>. \label{HL=ml}
\end{equation}
for the eigenvalues ${\mu}_{l(s)} = \frac{1}{2}(h_{11} + h_{22})\,
- (+)\,\frac{1}{2}[(h_{11} - h_{22})^{2} + 4h_{12} h_{21}]^{1/2}
\equiv m_{l(s)} - \frac{i}{2} {\gamma}_{l(s)}$, where $m_{l(s)},
{\gamma}_{l(s)}$ are real. These eigenvectors correspond to the
long (the vector $|l>$) and  short (the vector $|s>$) living
superpositions of $K_{0}$ and  $\overline{K_{0}}$.

Using the eigenvectors
\begin{equation}
|K_{1(2)}> \stackrel{\rm def}{=} 2^{-1/2} (|{\bf 1}> + (-) |{\bf
2}>) , \label{new8}
\end{equation}
of the CP--transformation for the eigenvalues $\pm  1$ (we define
${\cal C}{\cal P} |{\bf 1}> = - |{\bf 2}>$, \linebreak ${\cal
C}{\cal P} | {\bf 2}> = - |{\bf 1}>$), vectors  $|l>$ and $|s>$
can be expressed as follows
\begin{equation}
|l(s)> \equiv (1 + |{\varepsilon}_{l(s)}|^{2})^{- 1/2} [|K_{2(1)}
> + {\varepsilon}_{l(s)} |K_{1(2)} > ] . \label{r6}
\end{equation}
This last relation leads to the following formula for the product
$<s|l>$,
\begin{equation}
<s|l> \equiv N ({\varepsilon}_{s}^{\ast} +
{\varepsilon}_{l}^{\ast}), \label{s-l}
\end{equation}
where $N = N^{\ast} = [(1 + |{\varepsilon}_{s}|^{2}) (1 +
|{\varepsilon}_{l}|^{2})]^{- 1/2}$. By means of the following
parameters
\begin{equation}
\varepsilon \stackrel{\rm def}{=} \frac{1}{2} ( {\varepsilon}_{s}
+ {\varepsilon}_{l} ) , \label{r7}
\end{equation}
\begin{equation}
\delta \stackrel{\rm def}{=} \frac{1}{2} ( {\varepsilon}_{s} -
{\varepsilon}_{l} ), \label{r8}
\end{equation}
which are usually are used to describe the scale  of CP-- and
possible CPT -- violation effects \cite{4,5,Lee-qft,dafne,data},
product (\ref{s-l}) can be expressed as follows
\begin{equation}
<s|l> \equiv 2N (\Re \,{\varepsilon} -i \, \Im \, \delta ).
\label{s-l-1}
\end{equation}
There is
\begin{equation}
\delta \simeq  \frac{h_{11} - h_{22} }{2({\mu}_{s} - {\mu}_{l})}
\equiv {\delta}_{||} \, e^{ i {\phi}_{SW}} + {\delta}_{\perp} \,
e^{i ({\phi}_{SW} + \pi /2)}, \label{delta}
\end{equation}
in the case of $|{\varepsilon}_{s}| \ll 1 $ and
$|{\varepsilon}_{l}| \ll 1$ (see, eg. \cite{data}, p. 560). Here
${\phi}_{SW}$ is the superweak phase, $ \tan \, {\phi}_{SW} =
\frac{2(m_{l}- m{s})}{{\gamma}_{s} - {\gamma}_{l}}$, and
\begin{eqnarray}
{\delta}_{||} &=& \frac{1}{4} \frac{{\Gamma}_{11} -
{\Gamma}_{22}}{\sqrt{(m_{s} - m_{l})^{2} + \frac{1}{4}
({\gamma}_{s} - {\gamma}_{l})^{2}}}, \label{delta2}\\
{\delta}_{\perp} &=& \frac{1}{2} \frac{\Re \, (h_{11} -
h_{22})}{\sqrt{(m_{s} - m_{l})^{2} + \frac{1}{4} ({\gamma}_{s} -
{\gamma}_{l})^{2}}}, \label{delta3}
\end{eqnarray}
are the real parameters. Thus
\begin{equation}
\Im \, \delta = {\delta}_{||} \,\sin \, {\phi}_{SW} \; + \;
{\delta}_{\perp} \, \cos \, {\phi}_{SW} . \label{Im-delta}
\end{equation}

The consequence of (\ref{LOY-h=h}), (\ref{LOY-m=m}) is that in CPT
invariant but CP noninvariant system  ${\delta}_{||} =
{\delta}_{||}^{LOY} = 0$ and ${\delta}_{\perp} =
{\delta}_{\perp}^{LOY} = 0$ which leads to the standard result
$\Im {\delta}^{LOY} =0$ (here ${\delta}^{LOY}$ denotes the
parameter $\delta$, (\ref{delta}), calculated for $H_{||} =
H_{LOY}$). From this property and (\ref{s-l-1}) the conclusion
that the product $<s|l>$ must be real is drawn in the literature.
This conclusion is considered as the standard result. Note that in
the light of the main result of Sec. 2 and {\em Conclusion 1} such
a conclusion seems to be wrong in the case of the exact effective
Hamiltonian $H_{||}$, that is, in the case of the exact theory.
From {\em Conclusion 1} one infers that there must be
${\delta}_{\perp} \neq 0$, (\ref{delta3}) in the case of CPT
invariant but CP noninvariant system and therefore it must be $\Im
\, \delta \neq 0$ (see (\ref{Im-delta}) in such a system. This
means that the right hand side of the relation (\ref{s-l-1}) is a
complex number and therefore in the case of conserved CPT-- and
violated CP--symetries, in contrast with the standard result,
there must be $<s|l> \neq <s|l>^{\ast}$ in the real systems.

Properties of the real systems discussed above and described in
{\em Conclusion 1} are unobservable for the LOY approximation. In
order to obtain at least an estimation of the effects described in
these Conclusions, the matrix elements of $H_{\parallel}$ should
be calculated much more exactly than it is possible within the LOY
theory. A proposal of a more exact approximation is given in
\cite{9,10,11a,is}. This approximation is based on the
Krolikowski--Rzewuski equation for a distinguished component of
the state vector \cite{KR}. All CP -- and CPT -- transformation
properties of the effective Hamiltonian $H_{\parallel}$ calculated
within this approximation are consistent with similar  properties
of the exact effective Hamiltonian and with the result obtained in
this paper and \cite{plb2002} .

Within the mentioned more accurate approximation one finds for
diagonal matrix  elements  of $H_{\parallel} \simeq H_{||}^{(1)}
\stackrel{\rm def}{=} \lim_{t \rightarrow \infty} H_{||}^{(1)}(t)
$ that the CPT--invariant system in contradistinction to the
property (\ref{LOY-h=h}) obtained within the LOY theory
\cite{8,9,10,is}
\begin{equation}
h_{11}  \neq h_{22}, \label{b10}
\end{equation}
and
\begin{equation}
h_{11}(t=0) = H_{11} \equiv H_{22} = h_{22}(t=0). \label{t=0}
\end{equation}
The relation (\ref{t=0})  is consistent with the properties
(\ref{l8}) and (\ref{H=0})  and the result obtained in
\cite{plb2002}.

Assuming that
\begin{equation}
|H_{12}| \ll |H_{0} | , \label{b11}
\end{equation}
where $H_{0} \stackrel{\rm def}{=}\frac{1}{2} (H_{11} + H_{22})$,
( $H_{0} = H_{11} \equiv H_{22}$ if (\ref{[CPT,H]}) holds), one
finds within the mentioned more accurate approximation that (see
\cite{hep-ph/0202253} and (77) in \cite{10})
\begin{equation}
\Delta h \stackrel{\rm def}{=} h_{11} - h_{22} \simeq H_{12}
\frac{ \partial {\Sigma}_{21} (x) }{\partial x}
\begin{array}[t]{l} \vline \, \\ \vline \,
{\scriptstyle x = H_{0} } \end{array} - H_{21} \frac{ \partial
{\Sigma}_{12} (x) }{\partial x}
\begin{array}[t]{l} \vline \, \\ \vline \,
{\scriptstyle x = H_{0} }\end{array} \neq 0. \label{delta-h}
\end{equation}
Here
\begin{equation}
\Sigma ( \epsilon ) = PHQ \frac{1}{QHQ - \epsilon - i 0} QHP.
\label{r24}
\end{equation}
and ${\Sigma}_{jk} ( \epsilon ) = < {\bf j} \mid \Sigma ( \epsilon
) \mid {\bf k} >$.

From the result (\ref{delta-h}) it follows that $\Delta h = 0$ can
be achieved only if $H_{12}=H_{21} = 0$. Relation (\ref{new3})
implies that $H_{12} \equiv <{\bf 1}|H_{int}|{\bf 2}>$. If $|{\bf
1}> \equiv |K_{0}>$ and $|{\bf 2}> \equiv |{\overline{K}}_{0}>$
then the strangeness $S$ of the particle "1" equals $ S= +1$ while
that of "2" is $S= - 1$. Therefore the interpretation of the
hypothetical property $<{\bf 1}|H_{int}|{\bf 2}> \neq 0$, which
can be met in the literature, is that the first order $|\Delta S|
= 2$ transitions are allowed \cite{Jarlskog,Lee-qft,Bigi}.

So, the property $H_{12}=H_{21} = 0$ means that if the first order
$|\Delta S| = 2$ transitions are forbidden in the $K_{0},
{\overline{K}}_{0}$ complex then predictions following from the
use of the mentioned more accurate approximation  and from the LOY
theory should lead to the the same masses for $K_{0}$ and for
${\overline{K}}_{0}$. This does not contradict the {\em Conclusion
1} following from the results of Sec. 2 derived for the exact
$H_{||}$ or the rigorous result of \cite{plb2002}: the mass
difference is very, very  small and should arise at higher orders
of this more accurate approximation.

On the other hand from (\ref{delta-h}) it follows that in the
considered approximation  $\Delta h \neq 0$ if and only if $H_{12}
\equiv <{\bf 1}|H_{int}|{\bf 2}> \neq 0 $. This means that if
measurable deviations from the LOY predictions concerning the
equality of masses  of, e.g. $K_{0}, {\overline{K}}_{0}$ mesons
are ever detected in some tests, then the most plausible
interpretation of this result will be the existence of
interactions allowing the first order $|\Delta S| = 2$ transitions
in the system considered \cite{hep-ph/0202253}.

Within the use of the toy Fridrichs--Lee model \cite{chiu,9} the
following estimation was found  in \cite{plb2002}:
\begin{equation}
\frac{\Re \, (\Delta h )} {\Re \, (h_{11} +  h_{22}) } \sim 9,25
\times 10^{-18} \, (\Im \, (<{\bf 1}|H_{int}|{\bf 2}>)) \; [{\rm
MeV}]^{-1}. \label{hz-h0}
\end{equation}
This and the estimation $ \frac{|m_{K_{0}} -
m_{{\overline{K}}_{0}}|}{m_{K-average}}< 10^{-18}$ (where
$m_{\alpha}, (\alpha = K_{0}, {\overline{K}}_{0})$ denotes masses
of $K_{0}$ and ${\overline{K}}_{0}$--meson respectivelly), which
can be found in \cite{data}, show that possible deviations from
the standard picture, that is, from the LOY predictions are much
too small to be observed with the present experiments.

Confronting relations (\ref{LOY-h=h}) with (\ref{l18}), one should
remember that, in fact, $H_{LOY}$ can be considered as the lowest,
nontrivial order approximation in the perturbation $H^{(1)}$: All
the terms to higher orders than $(H^{(1)})^{2}$ are neglected in
$H_{LOY}$ \cite{Lee1} --- \cite{improved}. It is obvious that
CPT-- and other transformation properties of such an approximate
effective Hamiltonian and of the exact one need not be the same.
Taking into account all the above, it seems that for the proper
understanding of the CPT--invariance tests and CPT--invariance, or
possible CPT--violation phenomena it is necessary to consider
higher order contributions into the LOY--type effective
Hamiltonian than those contained in $H_{LOY}$ or to use a more
accurate approximation than LOY.

The result (\ref{LOY-h=h}) of the LOY approximation is model
independent whereas, within the mentioned  more accurate
approximation, the magnitude of \linebreak $\Re \,(h_{11} -
h_{22})$ depends on the model of interactions considered. So a new
possibility of the verification of models of weak interactions
arises.

It also seems, that above results have some meaning when attempts
to describe possible deviations from conventional quantum
mechanics are made and when possible experimental tests of such a
phenomenon and CPT--invariance in the neutral  kaons system are
considered \cite{12,13}. In such a case a very important role is
played by nonzero contributions to $(h_{11} - h_{22})$
\cite{12,13}: The correct description of these deviations and
experiments mentioned is impossible without taking into account
the results of this Section and the above Sec. 2. This can not be
performed within the LOY approach  and requires more exact
approximations. It seems that the approximation described and
exploited in \cite{8} --- \cite{10} may be a more effective tool
for  this  purpose.

The above considerations suggest that tests consisting of a
comparison of the equality  of  the  decay laws of ${\rm K}_{0}$
and ${\overline{\rm K}}_{0}$ mesons seem to be the only completely
model  independent tests  for  verifying  the CPT--invariance in
such and similar systems.

Taking into account all the above, it seems that all theories
describing the  time  evolution  of  the neutral kaon and similar
systems  by  means  of  the  effective Hamiltonian $H_{\parallel}$
governing their  time  evolution,  in which the CPT--invariance of
the total system  leads to the property (\ref{LOY-h=h}) for this
$H_{\parallel}$, (such as LOY theory \cite{Lee1} --- \cite{5}
based on  the  WW approximation), can not lay claim to being the
exact and correct  description of  all aspects of the effects
connected with the  violation  or nonviolation   of   the   CP--
and especially CPT--symmetries. (It occurs probably because of the
fact that such theories cannot exactly satisfy unitarity
\cite{kabir} and lead to inconsistencies of CPT--symmetry
properties of the $H_{\parallel}$ and the total Hamiltonians $H$
\cite{is}). Also, it seems that results of the experiments with
neutral kaons, etc., designed  and carried out on  the  basis of
expectations of  theories within the  WW approximation, such as
tests of CPT invariance (at least the results of  those in which
CPT--invariance or CPT--noninvariance of $H_{\parallel}$ generated
by such invariance  properties  of $H$ were essential), should be
revised using other methods  than the WW approach.

The most important observation which follows from the results of
Sec. 2 ({\em Conclusion 1}) and of \cite{plb2002} is the following
one:  In CPT invariant system Quantum Theory allows simultaneously
created at the instant $t_{0} = 0$ unstable particles and their
antiparticles as particles with the same masses to have slightly
different masses for $t > t_{0}$. Thus some matter--antimatter
asymmetry can arise in such system, which can have cosmological
consequences \cite{m-antim}.

\renewcommand{\theequation}%
{\Alph{section}.\arabic{equation}}
\appendix
\section{Appendix}
\setcounter{equation}{0}

The aim of this Appendix is to calculate the commutator $[ \Theta
,H_{\parallel}(t)]$ discussed in Sec.  2 and to  study  some of
its applications. In order to calculate this commutator it is
convenient to express $H_{\parallel}(t)$ by means of  the  formula
(\ref{l7a1}),  and then  to   use   assumption (\ref{[CPT,P]}),
the definition of $[U_{\parallel}(t)]^{-1}$ (\ref{l6}) and the
following property
\begin{equation}
P [U_{\parallel}(t)]^{-1} =  [U_{\parallel}(t)]^{-1} P \equiv P
[U_{\parallel}(t)]^{-1}P, \label{A1}
\end{equation}
which is the consequence of (\ref{l6}). This last observation
together with the property (\ref{l4}) means that the identity
(\ref{l5}) can be replaced by the following one:
\begin{equation}
H_{\parallel} \equiv H_{\parallel}(t)  \equiv i
\frac{{\partial}U_{\parallel}(t)}{\partial t}
[U_{\parallel}(t)]^{-1} P. \label{A2}
\end{equation}

Now one can consider a commutator $[\Theta, P
[U_{\parallel}(t)]^{-1} ]$. It  is the only  non\-tri\-vial
relation, necessary for the calculation of $[ \Theta
,H_{\parallel}(t)]$. Using property (\ref{A1}) and definition
(\ref{l6}),  we find (here the assumption (\ref{[CPT,P]}) is
crucial)
\begin{eqnarray}
[\Theta, P [U_{\parallel}(t)]^{-1} ] &=& \Theta P
[U_{\parallel}(t)]^{-1} - P [U_{\parallel}(t)]^{-1}
\Theta   \nonumber \\
&=& \Theta P [U_{\parallel}(t)]^{-1} - P [U_{\parallel}(t)]^{-1} P
\Theta  \nonumber \\
&=& P U_{\parallel}^{-1} \Big( U_{\parallel} \Theta -
\Theta U_{\parallel} \Big) U_{\parallel}^{-1} \label{A3} \\
&=& - P U_{\parallel}^{-1} [ \Theta , U_{\parallel} ]
U_{\parallel}^{-1}  \nonumber \\
& \equiv & - P U_{\parallel}^{-1} P [ \Theta , U ] P
U_{\parallel}^{-1}. \label{A4}
\end{eqnarray}

Properties (\ref{A1}) and expression (\ref{l7a1}) lead to the
following formulae
\begin{eqnarray}
[ \Theta , H_{\parallel}(t) ] &=& [ \Theta ,
PHUPU_{\parallel}^{-1} P]  \nonumber \\
&=& [ \Theta , PH ] UPU_{\parallel}^{-1} +
PH [ \Theta, UP U_{\parallel}^{-1} ] \nonumber \\
&=& P [ \Theta , H ] UP U_{\parallel}^{-1} \label{A5}  \\
&+& PH \Big{\{} [ \Theta , UP ] U_{\parallel}^{-1} + UP [ \Theta ,
P U_{\parallel}^{-1} ] \Big{\}}. \nonumber
\end{eqnarray}
All steps  in  the  above  formulae and in formulae leading to
(\ref{A2}) have  been  performed  without changing the order of
operators appearing in products of type ${\Theta} H, {\Theta}
U(t)$, etc.. By  virtue  of  the  assumption  (\ref{[CPT,P]})
only the order  of operators $\Theta$ and $P$ in products
${\Theta}P$, etc., can be changed when it is necessary.

Now, defining
\begin{equation}
{\cal A} (t)  \stackrel{\rm def}{=}  P [ \Theta , H ] UP
U_{\parallel}^{-1} , \label{A6}
\end{equation}
(which equals (\ref{l11}) ) and taking into account (\ref{A3}),
one can obtain formula (\ref{l10}) from (\ref{A4})
\begin{eqnarray*}
[ \Theta , H_{\parallel}(t) ] & \equiv & {\cal A}(t) + PH [ \Theta
, U ]
PU_{\parallel}^{-1}  \\
&+& PHUP \Big{\{} - U_{\parallel}^{-1} P [ \Theta, U ]
P U_{\parallel}^{-1} \Big{\}}  \\
& = & {\cal A}(t) + {\cal B}(t) ,
\end{eqnarray*}
where (see (\ref{l12}))
\[
{\cal B}(t) = \Big{\{} PH - PHUP U_{\parallel}^{-1} P \Big{\}} [
\Theta , U] PU_{\parallel}^{-1},
\]
or (by means of (\ref{l7a1}))
\[
{\cal B}(t) \equiv \Big{\{} PH - H_{\parallel} P\Big{\}} [ \Theta
, U] PU_{\parallel}^{-1},
\]
(i.e.,  simply  (\ref{l13})  ),  and   due   to   the   properties
(\ref{l7a}), (\ref{l7b})
\[
{\cal B}(t) = \Big{\{} PHQ - V_{\parallel} P\Big{\}} [ \Theta , U]
PU_{\parallel}^{-1} ,
\]
that is formula (\ref{l14}).

Let us now consider some details of the derivation of the relation
(\ref{l17}).  Taking into account the  properties  of  the
anti--unitary operator $\Theta$ and the CPT--trans\-for\-ma\-tion
properties of states $| K_{0}>, |{\overline K}_{0}> $, etc., (see
Sec. 2), without any assumptions about the commutator $[\Theta ,
H]$, one can transform the matrix element $<{\bf 2}|\Theta
H_{\parallel}(t) \Theta^{-1}|{\bf 2}>$ appearing in (\ref{l17}) as
follows
\begin{eqnarray*}
<{\bf 2}|\Theta H_{\parallel}(t) \Theta^{-1}|{\bf 2}> & \equiv &
<{\overline K}_{0}|\Theta H_{\parallel}(t)
\Theta^{-1}|{\overline K}_{0}> \\
& \equiv & <{\Theta} K_{0}, \Theta H_{\parallel}(t)
\Theta^{-1} {\Theta} K_{0}> \\
& = & <{\Theta}^{-1} \Theta H_{\parallel}(t) \Theta^{-1} K_{0},
{\Theta}^{-1} {\Theta} K_{0}> \\
& = & <H_{\parallel}(t) K_{0}, K_{0}> \\
& = & <K_{0}, H_{\parallel}(t) K_{0}>^{\ast} \\
& \equiv & <{\bf 1}| H_{\parallel}(t)|{\bf 1} >^{\ast} \equiv
h_{11}(t)^{\ast}.
\end{eqnarray*}
This last relation and the following consequence of ({\ref{l16})
\[
<{\bf 2}| \Theta H_{\parallel}(t) {\Theta}^{-1}| {\bf 2}> - <{\bf
2}| H_{\parallel}(t)|{\bf 2}> \equiv <{\bf 2}| ({\cal A}(t) +
{\cal B}(t)){\Theta}^{-1}| {\bf 2}>,
\]
yield
\begin{equation}
h_{11}(t)^{\ast} - h_{22}(t) = <{\bf 2}| \bigl( {\cal A}(t) +
{\cal B}(t) \bigr) {\Theta}^{-1}|{\bf 2}>,
\end{equation}
i.e., the formula (\ref{l17}).


\begin{thebibliography}{10}
\bibitem{Lee1} T. D. Lee, R. Oehme  and  C.  N.  Yang,  Phys.  Rev.,
{\bf 106}, (1957) 340.
\bibitem{Lee2} T. D. Lee and C. S.  Wu,  Annual  Review  of  Nuclear
Science, {\bf 16}, (1966) 471. Ed.:  M.  K.   Gaillard   and   M.
Nikolic,   Weak Interactions, (INPN et de Physique des Particules,
Paris,  1977); Chapt. 5, Appendix A. S. M. Bilenkij, Particles and
nucleus, vol.  1.  No  1 (Dubna 1970), p. 227 [in Russian]. P.  K.
Kabir,  The  CP-puzzle, Academic Press,  New York 1968.
\bibitem{4} J. W. Cronin, Rev. Mod. Phys. {\bf 53}, (1981) 373.
J. W. Cronin, Acta Phys. Polon., {\bf B15}, (1984) 419. V. V.
Barmin, et al., Nucl. Phys. {\bf B247}, (1984) 293. L. Lavoura,
Ann. Phys. (NY), {\bf 207}, (1991) 428. C. Buchanan, et al., Phys.
Rev. {\bf D45}, (1992) 4088. C. O. Dib, and R. D. Peccei, Phys.
Rev.,{\bf D46}, (1992) 2265. R. D. Peccei, CP and  CPT  Violation:
Status  and  Prospects, Preprint  UCLA/93/TEP/19,  University  of
California,  June 1993.
\bibitem{5} E. D. Comins and P. H. Bucksbaum, Weak interactions of
Leptons and Quarks, (Cambridge University Press, 1983). T. P.
Cheng and L. F. Li, Gauge Theory of Elementary Particle Physics,
(Clarendon, Oxford 1984).
\bibitem{Jarlskog}
K. Kleinkchnet, in: CP Violation, ed. C. Jarlskog, (World
Scientific, 1989), p.p. 41 --- 104.
\bibitem{Lee-qft}
T. D. Lee, Particle Physics and Introduction to Field Theory,
(Harwood Academic Publishers, 1990).
\bibitem{Bigi}
L. I.  Bigi and A. I. Sanda, CP Violation, (Cambridge University
Press, 2001).
\bibitem{6} Yu. V. Novozhilov, Introduction to the Theory
of Elementary Particles (Nauka, Moskow 1972), (in Russian). W. M.
Gibson and B. R. Pollard, Symmetry Principles in  Elementary
Particle  Physics,  (Cambridge  University  Press, 1976).
\bibitem{chiu} C. B. Chiu and E. C. G. Sudarshan, Phys. Rev.
{\bf D 42} (1990) 3712; E. C. G. Sudarshan, C. B. Chiu and G.
Bhamathi, Unstable Systems in Generalized Quantum Theory, Preprint
DOE-40757-023 and CPP-93-23, University of Texas, October 1993.
\bibitem{dafne} L. Maiani, in "The Second Da$\Phi$ne Physics Handbook",
vol. 1, Eds. L. Maiani, G. Pancheri and N. Paver, SIS ---
Pubblicazioni, INFN  --- LNF, Frascati, 1995; pp. 3 --- 26.
\bibitem{improved}
K. Urbanowski and J. Piskorski, Improved Lee, Oehme and Yang
approximation, Preprint of the Pedagogical University No WSP--IF
98--51, Zielona G\'{o}ra, March 1998: physics/9803030; Found.
Phys., {\bf 30}, (2000), 839.
\bibitem{horwitz} L. P. Horwitz and J. P. Marchand, Helv. Phys.
Acta {\bf 42} (1969) 801.
\bibitem{acta83} K. Urbanowski, Acta Physica Polonica, {\bf B 14},
(1983), 485.
\bibitem{11} K. Urbanowski, Phys. Lett. {\bf B 313},
(1993) 374.
\bibitem{data}
K. Hagiwara {\em et al}, Review of Particle Physics, Phys. Rev.,
{\bf D 66}, (2002), Part I, No 1--I.
\bibitem{plb2002} K. Urbanowski, Physics Letters {\bf B 540},
(2002), 89; hep--ph/0201272.
\bibitem{leonid1}
L. A. Khalfin, Preprints of the University of Texas at Austin: New
Results on the CP--violation problem,  (Report DOE--ER40200-211,
Feb. 1990); A new CP--violation effect and a new possibility for
investigation of $K_{S}^{0}, K_{L}^{0} (K^{0},
{\overline{K}}^{0})$ decay modes, (Report DOE--ER40200-247, Feb.
1991).
\bibitem{leonid2}
L. A. Khalfin, Foundations of Physics, {\bf 27}, (1997), 1549 and
references one can find therein.
\bibitem{kabir}P. K. Kabir and A. Pilaftsis, Phys. Rev. {\bf A 53},
(1996), 66.
\bibitem{8} K. Urbanowski, Int. J. Mod. Phys. {\bf A 7}, (1992)
6299. K. Urbanowski, Phys. Lett. {\bf A171}, (1992) 151.
\bibitem{9} K. Urbanowski, Int. J. Mod.  Phys.  {\bf  A 8},  (1993)
3721.
\bibitem{10} K. Urbanowski, Int. J. Mod.  Phys.  {\bf A 10}, (1995)
1151.
\bibitem{pra} K. Urbanowski, Phys. Rev. {\bf A 50}, (1994) 2847.
\bibitem{hep-ph/9803376}
K. Urbanowski, CPT transformation properties of the exact
effective Hamiltonian for neutral kaon and similar complexes,
hep--ph/9803376.
\bibitem{messiah} A. Messiah, Quantum Mechanics, vol. 2, (Wiley,
New York 1966).
\bibitem{bohm} A. Bohm, Quantum Mechanics: Foundations and
Applications, 2nd ed., (Springer, New York 1986).
\bibitem{15} E. P. Wigner, in: "Group Theoretical Concepts
and Methods in Elementary Particle Physics", ed.: F. G\"{o}resy,
(New York 1964).
\bibitem{16} J. S. Bell and J. Steinberger, in: "Oxford Int. Conf.
on Elementary Particles 19/25 September 1965:  Proceedings",  Eds.
T. R. Walsh, A. E. Taylor, R.  G.  Moorhouse  and  B.  Southworth,
(Rutheford High Energy Lab., Chilton, Didicot 1966), pp.  195  ---
222.
\bibitem{cpt} W. Pauli, in: "Niels Bohr and the Developmnet of
Physics". ed. W. Pauli (pergamon Press, London, 1955), pp. 30 ---
51. G. Luders, Ann. Phys. (NY) {\bf 2} (1957) 1. . R. Jost, Helv.
Phys. Acta {\bf 30} (1957) 409.R.F. Streater and A. S. Wightman,
"CPT, Spin, Statistics and All That" (Benjamin, New York, 1964).
N. N. Bogolubov, A. A. Logunov and I. T. Todorov, "Introduction to
Axiomatic Field Theory" (Benjamin, New York, 1975).
\bibitem{kobayashi}
M. Kobayashi and A. I. Sanda, Phys. Rev. Letters, {\bf 69},
(1992), 3139.
\bibitem{hep-ph/0202253}
K. Urbanowski, A new interpretation of one CPT violation test for
$K_{0} - {\overline{K}}_{0}$ system, hep--ph/0202253.
\bibitem{11a} K. Urbanowski, Int. J. Mod. Phys. {\bf A 7}, (1992)
6299.
\bibitem{is} K. Urbanowski,
Int. J. Mod. Phys. {\bf A 13}, (1998), 965.
\bibitem{KR} W. Krolikowski and J. Rzewuski, Bull. Acad. Polon.
Sci. {\bf 4} (1956) 19. W. Krolikowski and J. Rzewuski, Nuovo.
Cim. {\bf B 25} (1975) 739 and refernces therein.
\bibitem{12}
J. Ellis, J. S. Hagelin, D. V. Nanopoulos and M. Srednicki, Nucl.
Phys. {\bf B241}, (1984) 381. J. Ellis, N. E. Mavromatos and D. V.
Nanopoulos, Phys. Lett. {\bf B 293}, (1992) 142. J. Ellis, J. L.
Lopez, N. E. Mavromatos and D. V. Nanopoulos, Phys. Rev. {\bf D
53}, (1996) 3846.
\bibitem{13} P. Huet and M. E. Peskin, Nucl. Phys. {\bf B 434},
(1995) 3.  P. Huet, Testing Violation of CPT and Quantum Mechanics
in the $K_{0}-{\overline K}_{0}$ system, Preprint:
SLAC--Pub--6491, May 1994.
\bibitem{m-antim}
A. D. Sakharov, Pis'ma Zh. Eksp. Teor. Fiz., {\bf 5} (1967) 32 [in
Russian], [JETP Letters {\bf 5} (1967) 24]; G. R. Farrar and M. E.
Shaposhnikov, Phys. Rev. Lett. {\bf 70} (1993) 2833 and Phys. Rev.
{\bf D50} (1994) 774; M. B. Gavela, P. Hernandez, J. Orloff and O.
Pene, Nucl. Phys. {\bf B 430} (1994) 345, 382; K. A. Olive, Bing
Bang Baryogenesis, in {\em Proceedings of the 33rd International
Winter School on Nuclear and Particle Physics "Matter Under
Extreme Conditions"} --- Schladming (Austria) 1994, Eds. H. Latal
and W. Schweiger (Springer, Berlin, 1994), hep--ph/9404352; V. A.
Rubakov and M. E. Shaposhnikov, Uspekhi Fizicheskich Nauk, {\bf
166}, (1966), 493 and hep--ph/9603208; S. Sarkar, Reports on
Progres in Physics, {\bf 59}, (1996), 1493. M. Trodden, Rev. Mod.
Phys. {\bf 71},(1999), 1463; hep--ph/9803479. A. Riotto and M.
Trodden, Ann. Rev. Nucl. Part. Sci., {\bf 49},(1999), 35;
hep--ph/9901362. A. Riotto, Theories of baryogenesis, Preprint No
CERN--TH/98--204, hep--ph/9807454. A. D. Dolgov, Baryogenesis 30
years after, Preprint No TAC--1997--024
--- July 1997, hep--ph/9707419. A. D. Dolgov, Ya. B. Zeldovich,
Reviews of Modern Physics, {\bf 53}, (1981),1. O.Bertolami, Don
Colladay, V. A. Kostelecky and R. Potting, Phys. Lett. {\bf B
395}, (1997), 178. E. W. Kolb and M. S. Turner, {\em The early
Universe}, Addison--Wesley, 1993.
\end{thebibliography}
\end{document}